\documentclass[aps,prd,onecolumn,preprintnumbers,nofootinbib, 12pt]{revtex4-1}

\usepackage{amssymb,amsmath,amsthm,mathtools,physics}
\usepackage{slashed}
\usepackage{float}
\usepackage{multirow}
\allowdisplaybreaks

\usepackage[textsize=scriptsize]{todonotes}

\usepackage[linktoc=page]{hyperref}
\usepackage{cleveref}

\usepackage[usenames,dvipsnames]{xcolor}
\usepackage{subcaption}
\usepackage{graphicx}
\graphicspath{{./fig}}

\usepackage{soul}

\def\bar{\overline}

\def\vec{\vb*}

\newcommand{\bea}{\begin{eqnarray}}
\newcommand{\eea}{\end{eqnarray}}

\def\chf11#1{{}_{1}F_{1}\left( #1 \right)}

\def\cA{\mathcal{A}}
\def\cB{\mathcal{B}}
\def\cC{\mathcal{C}}
\def\cD{\mathcal{D}}
\def\cE{\mathcal{E}}
\def\cF{\mathcal{F}}

\def\cL{\mathcal{L}}

\def\cO{\mathcal{O}}
\def\cP{\mathcal{P}}

\begin{document}

%%%% 
\preprint{UCI-TR-2026-13, KIAS-Q26023}

\title{An Operator Analysis of Radiative Muon Decay}

\author{Arvind Rajaraman}
\email{arajaram@uci.edu}
\affiliation{Department of Physics and Astronomy, 
University of California, Irvine, CA 92697-4575, USA
}

\author{Chao-Hsiang Sheu}
\email{chsheu@kias.re.kr}
\affiliation{Quantum Universe Center, Korea Institute for Advanced Study, Seoul 02455, Korea
}

\begin{abstract}
 New physics beyond the Standard Model (SM) may induce new four-fermion interactions. We investigate the sensitivity of radiative muon decay, $\mu^-\to e^-\bar\nu_e\nu_\mu\gamma$, to such four-fermion interactions 
of mass dimension seven and eight.
For each operator, we calculate its interference with the SM amplitude and obtain an analytic expression for the decay rate.
We apply these to  a log-likelihood comparison for a benchmark sample of $10^5$ events, and find that these operator contributions can be distinguished from the SM prediction at characteristic coefficient magnitudes between $10^{-9}$ and $10^{-8}$. 
\end{abstract}

\maketitle
%%%%

\section{Introduction\label{sec:intro}}

Muon decay provides one of the cleanest low-energy laboratories for testing the charged-current weak interaction. Its lifetime determines the Fermi constant with high precision, while the energy and angular distributions of the decay products probe the Lorentz and chiral structure of the interaction \cite{Fetscher:1986uj,Kuno:1999jp,Webber:2010zf}. The agreement of these measurements with the Standard Model (SM) makes differential muon-decay observables sensitive probes of small new-physics effects.

{ The direct decay  $\mu^-\to e^-\bar\nu_e\nu_\mu$ strongly  constrains $G_F$. However, since neutrinos are typically unobservable, this decay only has one observable: the electron energy.}
The radiative channel $\mu^-\to e^-\bar\nu_e\nu_\mu\gamma$ adds a photon to the ordinary decay and therefore has more observables such as the  correlations among the electron energy $E_e$, photon energy $E_\gamma$, and their relative angle $\cos\theta_{e\gamma}$.

The  leading-order results for radiative muon decay were established in the classic analyses of Refs.~\cite{Kinoshita:1957zz,Fronsdal:1959zzb,ECKSTEIN1959297}. 
The Michel-parameter description originated with Michel's parametrization of the decay-electron spectrum \cite{Michel:1949,Michel:1950} and was extended to parity-violating and polarization observables by Bouchiat and Michel and by Kinoshita and Sirlin \cite{Bouchiat:1957,Kinoshita:1957zz,Kinoshita:1957polarization}. Radiative muon decay provides additional kinematic correlations that probe coupling combinations not separately accessible in the ordinary decay spectrum, including generalized Michel parameters absent from nonradiative leptonic decay distributions \cite{Arbuzov:2016ywn,Sehgal:2003radiative}.
On the SM side, the differential rate and branching fraction are known at next-to-leading order in QED with the full charged-lepton-mass dependence retained \cite{Fael:2015gua}. Experimentally, the MEG Collaboration measured the decay with a polarized muon sample and found a partial branching fraction consistent with the SM prediction \cite{Baldini:2013og}.

As the experimental results on radiative muon decay improve, these experiments  can potentially serve as a probe of new physics beyond the Standard Model.
Here we explore this possibility by studying an effective field theory with higher dimensional operators that can contribute to radiative muon decay. We find the leading contribution to muon decay, and explore the possibility that an experiment can bound or discover such operators. 
This provides an alternative perspective to other   model-independent studies of muon decay which parameterize new physics through momentum-independent scalar, vector, and tensor four-fermion couplings and organize their observable combinations as Michel parameters \cite{Fetscher:1986uj,Kuno:1999jp,Arbuzov:2016ywn}.

In \cref{sec:operators}, we construct the effective operators up to dimension 8 that can contribute to radiative muon decay.
These are listed below in \cref{tab:effop} in \cref{sec:operators}.

For each operator, we calculate the leading interference with the SM amplitude, integrate analytically over the unobserved neutrino phase space, and express the result as a triple-differential rate in $(E_e,E_\gamma,\cos\theta_{e\gamma})$. These results are presented in the following section (\ref{sec:rates}). These results  isolate the  linear response to one operator coefficient at a time and makes direct use of the kinematic information supplied by the photon.
We further show that the interference term probes the real components of certain coefficients ($d_{9}$ through $d_{14}$ in \cref{tab:effop}) and the imaginary components of other coefficients ($d_1,...,d_{8}$ in \cref{tab:effop}).  
This provides a framework to apply this effective field theory to any future experiment measuring radiative muon decay.

In 
\cref{{sec:likelihood}}, we use these results to  estimate the sensitivity of a future experiment to these new contributions to radiative muon decay.
We consider a   sample of $10^5$ radiative decays, binned in the kinematic variables $(E_e,E_\gamma,\cos\theta_{e\gamma})$. We further assume that there is exactly one nonzero operator beyond the Standard Model.  Assuming that the binned distribution follows the new physics, the distribution will deviate from the SM expectation. This can be parametrized by the log-likelihood of the two hypotheses; for a sufficiently large discrepancy in the log-likelihood, we can have a discovery of the new physics.

We show that with a required difference in log-likelihood of 5, we can probe coefficients of order $10^{-9}$ to $10^{-8}$.
The imaginary components can be probed at smaller coefficient magnitudes than the real components. The sensitivity also varies within each set, with $d_{7,8}$ giving the greatest reach among the imaginary components and $d_{11,12}$ among the real components. 

Conclusions and ideas for future work are presented at the end.

\section{Effective operators for radiative muon decay}
\label{sec:operators}

In this section, we systematically study effective operators up to dimension eight which contribute to radiative muon decays.

\begin{figure}[H]
  \centering 
  \begin{subfigure}[b]{0.46\textwidth}
    \includegraphics[width=\linewidth]{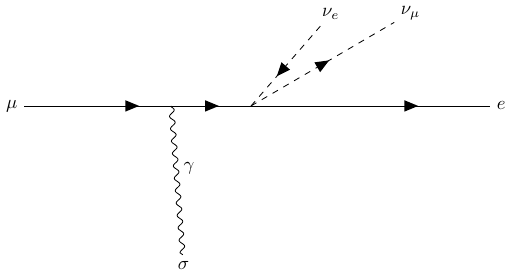}
  \end{subfigure}
  \hfill
  \begin{subfigure}[b]{0.46\textwidth}
    \includegraphics[width=\linewidth]{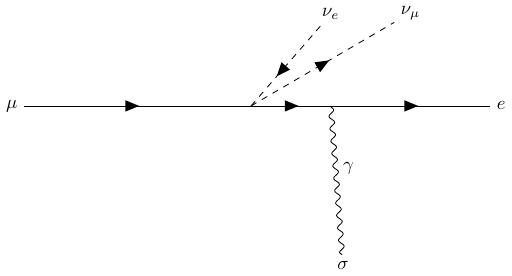}
  \end{subfigure}
  \caption{\small Tree-level SM processes of radiative muon decay}
  \label{fig:SMtree}
\end{figure}
We first note that in the Standard Model, the Fermi interaction that mediates muon decay is described by \cite{Czarnecki:2011mx}
\begin{align}
\cL_{F} = 2\sqrt{2}G_F\left( \bar{\psi}_{e}\gamma^{\rho}P_L\psi_{\mu} \right)\left( \bar{\psi}_{\nu_{\mu}}\gamma_{\rho}P_L\psi_{\nu_e} \right) 
+ {\rm h.c.}
\end{align}
where $G_F$ is the Fermi constant. Radiative muon decay occurs in  the SM at  tree-level, through the diagrams in \cref{fig:SMtree}.

Similarly, the new physics operators can also be written (using a Fierz transformation if necessary) as a product of a neutrino bilinear and a charged fermion bilinear. 
Further, since we are considering radiative muon decay, the operators can also include a photon.
An operator with no photons can still yield a photon after attaching an extra leg (similar to the SM operator) as shown in \cref{fig:SMtree}.

The full list of relevant operators up to dimension 8 is summarized in \cref{tab:effop} (the dimension-6 operator would correct $G_F$, and as such is heavily constrained. We have therefore dropped this operator). Here $D_{a}$ is the  covariant derivative, $F_{AB}$ is the field strength, and multi-indices $\gamma$-matrices are defined as
\begin{align}
\gamma^{ab} \equiv \frac{1}{2!}\gamma^{[a}\gamma^{b]}
\qand
\gamma^{abc} \equiv \frac{1}{3!}\gamma^{[a}\gamma^{b}\gamma^{c]}
\,
\end{align}
We have eliminated any terms with derivatives on $\bar{\psi}_{\bar{\nu}}$ by an integration by parts.
\begin{table}[t]
\centering
\renewcommand{\arraystretch}{1.6}
\makebox[0.95\linewidth][c]{%
\begin{tabular}{|c||c|}
\hline
~Dimension~ & \parbox{0.78\linewidth}{\centering Operators} \\
\hline\hline
\multirow{2}{*}{$7$} & \parbox{0.78\linewidth}{\centering
$\displaystyle\frac{1}{m_{\mu}^3}\;\bar{\psi}_{\bar{\nu}}
\gamma^{\rho}P_L\psi_{\nu}\times \bar{\psi}_{\bar{\mu}}
(d_1+d_2\gamma^5)D_{\rho}\psi_{e}$} \\
& \parbox{0.78\linewidth}{\centering
$\displaystyle\frac{1}{m_{\mu}^3}\;\bar{\psi}_{\bar{\nu}}
\gamma^{\rho}P_LD_a\psi_{\nu}\times \bar{\psi}_{\bar{\mu}}
\gamma^{\rho a}(d_{3}+d_{4}\gamma^5)\psi_{e}$} \\
\hline
\multirow{5}{*}{$8$} & \parbox{0.78\linewidth}{\centering
$\displaystyle\frac{1}{m_{\mu}^4}\;\bar{\psi}_{\bar{\nu}}
\gamma^{\rho}P_L\psi_{\nu} F_{B\rho} \times \bar{\psi}_{\bar{\mu}}
\gamma^{B}(d_5+d_6\gamma^5)\psi_{e}$} \\
& \parbox{0.78\linewidth}{\centering
$\displaystyle\frac{1}{m_{\mu}^4}\;
\bar{\psi}_{\bar{\nu}}
\gamma_{\rho}P_L\psi_{\nu}F_{ab}\times \bar{\psi}_{\bar{\mu}}
\gamma^{\rho ab}(d_{7}+d_{8}\gamma^5)\psi_{e}
$\rule{0pt}{2.4em}} \\
& \parbox{0.78\linewidth}{\centering
$\displaystyle\frac{1}{m_{\mu}^4}\;\bar{\psi}_{\bar{\nu}}
\gamma^{\rho}P_LD^a\psi_{\nu}\times \bar{\psi}_{\bar{\mu}}
\gamma_{\rho}(d_{9}+d_{10}\gamma^5)D_a\psi_{e}$\rule{0pt}{2.4em}} \\
& \parbox{0.78\linewidth}{\centering
$\displaystyle\frac{1}{m_{\mu}^4}\;
\bar{\psi}_{\bar{\nu}}
\gamma^{\rho}P_LD_a\psi_{\nu}\times \bar{\psi}_{\bar{\mu}}
\gamma^{a}(d_{11}+d_{12}\gamma^5)D_{\rho}\psi_{e}
$\rule{0pt}{2.4em}} \\
& \parbox{0.78\linewidth}{\centering
$\displaystyle\frac{1}{m_{\mu}^4}\;
\partial_a\bar{\psi}_{\bar{\nu}}
\gamma^{\rho}P_L\partial^a\psi_{\nu}\times \bar{\psi}_{\bar{\mu}}
\gamma_{\rho}(d_{13}+d_{14}\gamma^5)\psi_{e}
$\rule{0pt}{2.4em}} \\
\hline
\end{tabular}
}
\caption{\small Effective operators by mass dimension.}
\label{tab:effop}
\end{table}

\section{Decay rates}\label{sec:rates}

\subsection{Overview}
We now find the contribution to the decay rate from each operator.

  We focus on the   interference term between the SM operator and the new physics operators. 
 We write  the cross section as 
\begin{align}
\sigma \sim \left( \cA_{\rm SM} + \cA_{\rm new} \right) \overline{\left( \cA_{\rm SM} + \cA_{\rm new} \right)}
= \abs{\cA_{\rm SM}}^2 + \left( \cA_{\rm SM}\overline{\cA}_{\rm new} + {\rm h.c.} \right) + \cdots
\end{align}
where we assume tat the square of the new physics process is small and can be neglected, and we keep only contributions from the interference  term. This implies that  the effects are linear in the new physics coefficients, and hence we can consider each operator  separately.

The new physics operators are a product of a neutrino bilinear and a charged fermion bilinear. Since the neutrinos are typically unobservable, we can sum over the neutrino spins and integrate over the neutrino momenta. The integral over the neutrino phase space will therefore provide a factor in the final cross section depending only on the sum of the neutrino momenta $Q$.  
We will further assume that the electon spin is {\it not} measured in the experiment, and that we can only observe the sum over electron spins.

For the charged fermion factor, the calculation depends on the form of the operator. An operator which   has a photon  directly produces the corresponding vertex. For the operators with no photons, the photon emission can still  be generated by a diagram of the form
(\ref{fig:SMtree}), 
 in which case the relevant amplitude is of the form 
\begin{align}\label{eq:nogleg}
\cA^{\alpha}(\cO) = \cO(p_e)\frac{i}{(p_{m}-p_{\gamma})\cdot\gamma-m_{\mu}}(ie\gamma^{\alpha})
+ (ie\gamma^{\alpha})\frac{i}{(p_e+p_{\gamma})\cdot\gamma-m_e}\cO(p_e+p_{\gamma})
\end{align}

\subsection{The Standard Model}
The vertex in the Standard Model is a four fermion operator with no photon. The radiative decay  rate is of the form
\begin{align} \frac{\dd\Gamma^{(SM)}}{\dd{E_e}\dd{E_\gamma}\dd{\cos{\theta_{e\gamma}}}}\equiv \left( \frac{\abs{\vec{p}_e}E_{\gamma}}{256\pi^6 m_{\mu}} \right) \cdot (2\sqrt{2}G_F)^2 R^{(SM)}
\end{align}
where  the observables are the the electron energy $E_e$, photon energy $E_\gamma$, and their relative angle $\cos\theta_{e\gamma}$. We use $Q\equiv p_m-p_e-p_\gamma$ for the total neutrino momentum.

Furthermore the final factor can be written as a product
\bea
R^{(SM)}=I^{(1)}_{\alpha\beta}(Q)R^{\alpha\beta, SM} 
\eea
where 
\bea
I^{(1)}_{\alpha\beta}(Q) &\equiv -\frac{2\pi}{3}\left( Q^2\eta_{\alpha\beta} - Q_{\alpha}Q_{\beta} \right)
\eea
depends only on the sum of neutrino momenta, and 
the charged fermion line contributes a factor
\begin{align}
R^{\alpha\beta, SM} &= -\frac{1}{2}\eta_{\lambda\rho}\tr\left[ \cA^{\lambda}\left( \gamma^{\alpha}P_L \right)\left( \slashed{p}_{m}+m_{\mu} \right)\bar{\cA^{\rho}\left( \gamma^{\beta}P_L \right)}\left( \slashed{p}_e+m_e \right) \right]
\end{align}
For the explicit result, we use the following variables
\begin{align}
a&\equiv p_m\cdot p_\gamma
\,,
\qquad
b\equiv p_e\cdot p_\gamma
\,,
\qquad
t\equiv p_m\cdot p_e=m_\mu E_e
\,.
\end{align}
The fully contracted SM factor can then be organized as
\begin{align}
R^{(SM)}
={}&\frac{\pi e^2}{3}
\left[
\cF_{\rm SM}^{(0)}
+\left(\frac{m_e}{m_\mu}\right)^2\cF_{\rm SM}^{(1)}
+\left(\frac{m_e}{m_\mu}\right)^4\cF_{\rm SM}^{(2)}
\right]
\,,
\end{align}
where
\begin{align}
\cF_{\rm SM}^{(0)}
\equiv{}&
-\frac{(4a-3m_\mu^2+4t)(a^2+2at+2t^2)}{ab}
\nonumber\\[2mm]
&+\frac{1}{a^2}\left(
4a^3+2a^2m_\mu^2+16a^2t-3am_\mu^4
+2am_\mu^2t+16at^2
-3m_\mu^4t+4m_\mu^2t^2
\right)
\nonumber\\[2mm]
&-\frac{b}{a^2}\left(
4a^2+5am_\mu^2+12at-3m_\mu^4+8m_\mu^2t
\right)
+\frac{4b^2}{a^2}(a+m_\mu^2)
\,,
\\[2mm]
\cF_{\rm SM}^{(1)}
\equiv{}&
m_\mu^2\left[
\frac{(a+t)(4a-3m_\mu^2+4t)}{b^2}
+\frac{-5a^2+3am_\mu^2-2at-4m_\mu^2t+6t^2}{ab}
\right.
\nonumber\\[2mm]
&\left.
+\frac{2a^2-3am_\mu^2-6at+2m_\mu^4-3m_\mu^2t}{a^2}
+\frac{3b(a+m_\mu^2)}{a^2}
\right]
\,,
\\[2mm]
\cF_{\rm SM}^{(2)}
\equiv{}&
m_\mu^4\left[
-\frac{3a-2m_\mu^2+3t}{b^2}
+\frac{3}{b}
\right]
\,.
\end{align}

 This process is well studied in the literature and can be found in, for example, \cite{Kinoshita:1957zz,Fronsdal:1959zzb,ECKSTEIN1959297,Kuno:1999jp,Arbuzov:2016ywn}.

\subsection{The new physics operators}
The new physics operators separate into 
the following five classes depending on the existence of a photon  and their neutrino operator. 

The operators with an external photon are
\begin{align}
\cO^{(2)}_{\rho\tau} \equiv&~  i\Biggl[((p_\gamma)_B \eta_{\rho\tau}-(p_\gamma)_\rho \eta_{B\tau})
\gamma^{B}\frac{d_{5}+d_{6}\gamma^5}{m_{\mu}^4}
+e\eta_{\rho\tau}\frac{d_{1}+d_{2}\gamma^5}{m_{\mu}^3}
\nonumber\\
&\hspace{50mm}+\gamma_{\rho}^{~ab}((p_\gamma)_a \eta_{b\tau}-(p_\gamma)_b \eta_{a\tau}) \frac{d_{7}+d_{8}\gamma^5}{m_{\mu}^4}
\Biggr] 
\\[2mm]
\cO^{(4)}_{\rho\tau a} \equiv&~ ie\left( \gamma_{\rho}\frac{d_{9}+d_{10}\gamma^5}{m_{\mu}^4}\eta_{a\tau}+\gamma_{a}\frac{d_{11}+d_{12}\gamma^5}{m_{\mu}^4}\eta_{\rho\tau} \right)
\end{align}
which are separated according to the form of the neutrino operator.
The corresponding differential decay rates are
\begin{align} \frac{\dd\Gamma^{(j)}}{\dd{E_e}\dd{E_\gamma}\dd{\cos{\theta_{e\gamma}}}}\equiv \left( \frac{\abs{\vec{p}_e}E_{\gamma}}{256\pi^6 m_{\mu}} \right) \cdot 2\sqrt{2}G_F R^{(j)}\label{eq:rate}
\end{align}
with
\begin{align}
\label{eq:R2}
R^{(2)} =&~ \eta_{\sigma\tau}I^{(1)}_{\alpha\beta}\tr\left[ \cO^{(2),\alpha\tau}\left( \slashed{p}_{\rm muon}+m_{\mu} \right)\bar{\cA^{\sigma}(\cO_{\rm SM}^{\beta})}(\slashed{p}_e+m_e) \right] + {h.c.}
\\[2mm]
\label{eq:R4}
R^{(4)} =&~ \eta_{\sigma\tau}I^{(2)}_{\beta\rho\alpha}\tr\left[ \cO^{(4),\rho\tau\alpha}\left( \slashed{p}_{\rm muon}+m_{\mu} \right)\bar{\cA^{\sigma}(\cO_{\rm SM}^{\beta})}(\slashed{p}_e+m_e) \right] + {h.c.}
\end{align}

The operators with no external photon are
\begin{align}
\cO^{(1)}_{\rho} \equiv&~  
  \frac{1}{m_{\mu}^3}(d_{1}+d_{2}\gamma^5)(ip_{e})_{\rho} 
\\[2mm]
\cO^{(3)}_{\rho\tau} \equiv&~ \gamma_{\rho\tau}\frac{d_{3}+d_{4}\gamma^5}{m_{\mu}^3}+
\gamma_{\rho}\frac{d_{9}+d_{10}\gamma^5}{m_{\mu}^4}(ip_e)_\tau+
\gamma_{\tau}\frac{d_{11}+d_{12}\gamma^5}{m_{\mu}^4}(ip_e)_{\rho}
\\[2mm]
\cO^{(5)}_{\rho} \equiv&~ \frac{1}{m_{\mu}^4}\gamma_{\rho}(d_{13}+d_{14}\gamma^5)
\end{align}

and have decay rates of the form (\ref{eq:rate}) with 
\begin{align}\label{eq:R1}
R^{(1)} =&~\eta_{\sigma\tau}I^{(1)}_{\alpha\beta}\tr\left[ \cA^{\tau}(\cO^{(1),\alpha})\left( \slashed{p}_{\rm muon}+m_{\mu} \right)\bar{\cA^{\sigma}(\cO_{\rm SM}^{\beta})}(\slashed{p}_e+m_e) \right] + {h.c.}
\\[2mm]
\label{eq:R3}
R^{(3)} =&~ \eta_{\sigma\tau}I^{(2)}_{\beta\rho\alpha}\tr\left[ \cA^{\tau}(\cO^{(3),\rho\alpha})\left( \slashed{p}_{\rm muon}+m_{\mu} \right)\bar{\cA^{\sigma}(\cO_{\rm SM}^{\beta})}(\slashed{p}_e+m_e) \right] + {h.c.}
\\[2mm]
\label{eq:R5}
R^{(5)} =&~ \eta_{\sigma\tau}I^{(3)}_{\alpha\beta}\tr\left[ \cA^{\tau}(\cO^{(5),\alpha})\left( \slashed{p}_{\rm muon}+m_{\mu} \right)\bar{\cA^{\sigma}(\cO_{\rm SM}^{\beta})}(\slashed{p}_e+m_e) \right] + {h.c.}
\end{align}

where $\bar{\cA^{a}(\cO)} \equiv \gamma^{0}\cA^{a}(\cO)^{\dagger}\gamma^0$.

In these formulae, we have found factors from the integral over the neutrino sectors which are 
\begin{align}
I^{(2)}_{\alpha\beta\delta}(Q) \equiv -i\frac{\pi}{3}Q_{\delta}\left( Q^{2}\eta_{\alpha\beta}-Q_{\alpha}Q_{\beta} \right)
\,,\qquad
I^{(3)}_{\alpha\beta}(Q) \equiv \frac{\pi}{3}Q^2\left( Q^{2}\eta_{\alpha\beta}-Q_{\alpha}Q_{\beta} \right)
\,.
\end{align}

In the fermion line, the terms with $\epsilon^{\mu\nu\rho\sigma}$ vanish because there are only three independent momenta ($p_m, p_e, p_\gamma, Q$ must sum to zero).
The amplitude then only depends on dot products of momenta. To present the  expressions, it is convenient to also define
\begin{align}
s&\equiv Q^2=m_\mu^2-2t-2a+2b
\,.
\end{align}

Since the soft region is often not detected in the measurement, it suffices to consider the  rates in the limit $m_e\to0$. 

\subsection{Results}
For  $d_{9}$ through $d_{14}$, we find
\begin{align}
R_{9}
={}&\frac{\pi e^2 }{3m_\mu^4}\cF_{9}\times {\rm Re~}d_{9}
\,,
\qquad
R_{10}
=-\frac{\pi e^2}{3m_\mu^4}\cF_{9}\times {\rm Re~}d_{10}
\,,
\nonumber\\[2mm]
R_{11}
={}&\frac{\pi e^2}{3m_\mu^4}\cF_{11}\times {\rm Re~}d_{11}
\,,
\qquad
R_{12}
=-\frac{\pi e^2}{3m_\mu^4}\cF_{11}\times {\rm Re~}d_{12}
\,,
\nonumber\\[2mm]
R_{13}
={}&\frac{2\pi e^2s }{3m_\mu^4}\cF_{13}
\times {\rm Re~}d_{13}
\,,
\qquad
R_{14}
=-\frac{2\pi e^2s }{3m_\mu^4}\cF_{13}\times {\rm Re~}d_{14}
\,.
\end{align}

Here we have defined
\begin{align}
\cF_{9}
\equiv{}&
\frac{2(t-b)\cP_A}{a^2}
+\frac{(2t+a-3b)\cP_B}{ab}
+\frac{2(t+a-2b)\cP_C}{b^2}
-\frac{\cD_{9}}{a}
-\frac{\cE_{9}}{b}
\,,
\nonumber\\[2mm]
\cF_{11}
\equiv{}&
\frac{\cA_{11}}{a^2}
+\frac{\cB_{11}}{ab}
+\frac{\cC_{11}}{b^2}
+\frac{\cD_{11}}{a}
+\frac{\cE_{11}}{b}
\,,
\nonumber\\[2mm]
\cF_{13}
\equiv{}&
\frac{\cP_A}{a^2}
+\frac{\cP_B}{ab}
+\frac{\cP_C}{b^2}
\end{align}

The polynomials entering $\cF_{9}$ and $\cF_{13}$ are
\begin{align}
\cP_A={}&2\left(
3m_\mu^4b-3m_\mu^4t-5m_\mu^2ab+6m_\mu^2at
+4m_\mu^2b^2-8m_\mu^2bt+4m_\mu^2t^2
\right.
\nonumber\\
&\left.
\quad-4a^2b+2a^2t+4ab^2-4abt
\right)
\,,
\\[2mm]
\cP_B={}&-2\left(
3m_\mu^4b-4m_\mu^2ab-6m_\mu^2at+4m_\mu^2bt
-6m_\mu^2t^2+8a^2t-12abt
\right.
\nonumber\\
&\left.
\quad+16at^2+8b^2t-16bt^2+8t^3
\right)
\,,
\\[2mm]
\cP_C={}&2b\left(
3m_\mu^2a-2m_\mu^2b-4a^2+4ab-4at+2bt
\right)
\,,
\\[2mm]
\cD_{9}={}&2\left(
-5m_\mu^4b+6m_\mu^4t+12m_\mu^2ab-16m_\mu^2at
-6m_\mu^2b^2+20m_\mu^2bt-14m_\mu^2t^2
\right.
\nonumber\\
&\left.
\quad-4a^2b+8a^2t-16abt+16at^2+8b^2t-16bt^2+8t^3
\right)
\,,
\\[2mm]
\cE_{9}={}&2\left(
3m_\mu^4b-4m_\mu^2ab-6m_\mu^2at-2m_\mu^2b^2
+6m_\mu^2bt-6m_\mu^2t^2+8a^2t
\right.
\nonumber\\
&\left.
\quad+4ab^2-16abt+16at^2+8b^2t-16bt^2+8t^3
\right)
\,.
\end{align}
Note that the massless expressions containing $1/b$ are understood away from the strictly collinear endpoint $b=0$.

The remaining polynomials entering $\cF_{11}$ are
\begin{align}
\cA_{11}={}&4(t-b)\left(
m_\mu^4b-m_\mu^4t-3m_\mu^2ab+2m_\mu^2at+2a^2t
\right)
\,,
\\[2mm]
\cB_{11}={}&2\left(
-m_\mu^4ab+7m_\mu^4b^2-4m_\mu^4bt+2m_\mu^2a^2t
-4m_\mu^2ab^2-10m_\mu^2abt
\right.
\nonumber\\
&\left.
\quad+6m_\mu^2at^2-4m_\mu^2b^3+8m_\mu^2b^2t
-10m_\mu^2bt^2+4m_\mu^2t^3+4a^2bt+4ab^2t
\right)
\,,
\\[2mm]
\cC_{11}={}&-4b\left(
m_\mu^4b-m_\mu^2a^2+2m_\mu^2ab-m_\mu^2at
-2m_\mu^2b^2-2m_\mu^2bt+2abt+2bt^2
\right)
\,,
\\[2mm]
\cD_{11}={}&-2\left(
-3m_\mu^4b+12m_\mu^2ab-2m_\mu^2b^2+2m_\mu^2t^2
-12a^2b+8ab^2-8abt
\right)
\,,
\\[2mm]
\cE_{11}={}&2\left(
m_\mu^4b-8m_\mu^2ab+2m_\mu^2at+10m_\mu^2b^2
-10m_\mu^2bt+2m_\mu^2t^2
\right.
\nonumber\\
&\left.
\quad+8a^2b-12ab^2+8abt
\right)
\,.
\end{align}

For $  d_i$ for $i=1,...,8$
we have
\begin{align}
R_1
={}&\frac{4\pi e^2}{3m_\mu^3}\cF_{1}\times {\rm Im~}d_{1}
\,,
\qquad
R_2
=-\frac{4\pi e^2}{3m_\mu^3}\cF_{1}
\times {\rm Im~}d_{2}
\,,
\nonumber\\[2mm]
R_3
={}&\frac{4\pi e^2s }{3m_\mu^3}\cF_{3}
\times {\rm Im~}d_{3}
\,,
\qquad
R_{4}
=-\frac{4\pi e^2s }{3m_\mu^3}\cF_{3}
\times {\rm Im~}d_{4}
\,,
\nonumber\\[2mm]
R_5
={}&\frac{8\pi e }{3m_\mu^4}\cF_{5}
\times {\rm Im~}d_{5}
\,,
\qquad
R_6
=-\frac{8\pi e }{3m_\mu^4}\cF_{5}\times {\rm Im~}d_{6}
\,,
\nonumber\\[2mm]
R_{7}
={}&-\frac{16\pi e} {3m_\mu^4}\cF_{7}
\times {\rm Im~}d_{7}
\,,
\qquad
R_{8}
=\frac{16\pi e }{3m_\mu^4}\cF_{7}\times {\rm Im~}d_{8}
\,.
\end{align}

where we define
\begin{align}
\cF_{1}
\equiv{}&
\frac{\cA_{1}}{a^2}
+\frac{\cB_{1}}{b^2}
+\frac{\cC_{1}}{ab}
+\frac{\cD_{1}}{a}
+\frac{\cE_{1}}{b}
\,,
\nonumber\\[2mm]
\cF_{3}
\equiv{}&
\frac{\cA_{3}}{a^2}
+\frac{\cB_{3}}{b^2}
+\frac{\cC_{3}}{ab}
\,,
\nonumber\\[2mm]
\cF_{5}
\equiv{}&
\frac{\cD_{5}}{a}
+\frac{\cE_{5}}{b}
\,,
\nonumber\\[2mm]
\cF_{7}
\equiv{}&
\frac{\cD_{7}}{a}
+\frac{\cE_{7}}{b}
\,.
\end{align}

The nonvanishing numerator polynomials are
\begin{align}
\cA_{1}={}&
-2m_\mu(a-m_\mu^2)(t-b)^2
\,,
\\[2mm]
\cB_{1}={}&
-2m_\mu b\left(
-m_\mu^2b+a^2-ab+at+bt
\right)
\,,
\\[2mm]
\cC_{1}={}&
-m_\mu\left(
-m_\mu^2ab+5m_\mu^2b^2-4m_\mu^2bt+2a^2t-2ab^2
-6abt
\right.
\nonumber\\
&\left.
\quad+6at^2+4b^2t-8bt^2+4t^3
\right)
\,,
\\[2mm]
\cD_{1}={}&
m_\mu\left(
-3m_\mu^2b+6ab-4b^2+2bt+2t^2
\right)
\,,
\\[2mm]
\cE_{1}={}&
-m_\mu\left(
m_\mu^2b-4ab+2at+6b^2-8bt+2t^2
\right)
\,,
\\[2mm]
\cA_{3}={}&
3m_\mu(a-m_\mu^2)(t-b)
\,,
\\[2mm]
\cB_{3}={}&
3m_\mu b(a-b)
\,,
\\[2mm]
\cC_{3}={}&
m_\mu\left(
-3m_\mu^2b+ab+6at+2b^2-9bt+6t^2
\right)
\,,
\\[2mm]
\cD_{5}={}&
m_\mu^4b-3m_\mu^2ab-m_\mu^2at-m_\mu^2b^2
+6a^2b-2a^2t-4ab^2+6abt
\,,
\\[2mm]
\cE_{5}={}&
b\left(
-3m_\mu^2a+3m_\mu^2b-m_\mu^2t+4a^2-6ab+6at-2bt
\right)
\,,
\\[2mm]
\cD_{7}={}&
2m_\mu^4b-3m_\mu^2ab-2m_\mu^2at+3m_\mu^2b^2
-4m_\mu^2bt
\nonumber\\
&-4a^2b+6a^2t+4ab^2-6abt+4at^2
\,,
\\[2mm]
\cE_{7}={}&
b\left(
-3m_\mu^2a+3m_\mu^2b-2m_\mu^2t+4a^2-4ab
+6at-6bt+4t^2
\right)
\,.
\end{align}

The results above provide a complete characterization of radiative muon 
deacay up to dimension 8 operators.

\section{Future Experimental Reach And likelihood analysis}
\label{sec:likelihood}
In this section, we will show one possible application of our results above. Specifically, we use the results above to  how future experiments can bound, and possibly discover, new physics in muon decays.

We shall be agnostic about the nature of the new experiments. We shall assume that a certain number $N_\mu$ of muon radiative decays are identified (below we shall take this number to be $10^5$), and that the energies of the electron and photon are measured to some accuracy (below we shall take this number to be $20\%$). We then consider a binned sample of events distributed according to a theory which is the SM extended by one of the operators in \cref{tab:effop} with some value of the Wilson coefficient.  We can then calculate the likelihood of obtaining these results under the Standard Model hypothesis. This likelihood is related to the Wilson coefficient, and we can calculate the value of the Wilson coefficient that would yield a particular likelihood. This provides an estimate of the reach of the experiment.
We note that since the interference terms are linear in the Wilson coefficients, we can analyze each operator separately.

We work in the muon rest frame, and
 choose the parametrization
\begin{align}
p_{m} = \left( m_{\mu} , \vec{0} \right)
\,,\quad
p_{e} = \left( E_{e},0,0, \sqrt{E_e^2 - m_e^2} \right)
\,,\quad
p_{\gamma} = \left( E_{\gamma}, 0, E_{\gamma}\sin{\theta_{e\gamma}}, E_{\gamma}\cos{\theta_{e\gamma}} \right)
\end{align}
where $E_{e}, E_{\gamma}$, and $\theta_{e\gamma}$ are the energy of electron and photon, and the angle between the outgoing electron and the outgoing photon, respectively. 
A single radiative event is fully characterized, once the invisible neutrino pair is integrated out, by the observables $\vec{x}\equiv(E_e,E_\gamma,\cos\theta_{e\gamma})$. 
 The physical regime satisfies
\begin{align}\label{eq:physgrid}
m_e \leq E_{e} \leq \frac{m_{\mu}^2+m_e^{2}}{2m_{\mu}} 
\,,\qquad
0 \leq E_{\gamma} \leq  \frac{m_{\mu}^2+m_e^2-2m_{\mu}E_e}{2(m_{\mu}-E_e+\sqrt{E_e^2-m_e^2}\cos{\theta})}
\end{align}

The resolution of the experiment is modelled by 
discretizing the kinematically allowed region. We divide the range of each kinematical observable into $N$ equal bins (where $N$ is specified below to be 5), producing $N^{3}$ possible experimental results.
We label these possible results by a multi-index $b=(i_1,i_2,i_3)$ with $i_k=1,\dots,N$.

The total differential rate found above is  of the form
\begin{align}\label{eq:linrate}
\rho(\vec{x};d_i)
&\equiv \frac{\dd\Gamma}{\dd{E_e}\dd{E_\gamma}\dd{\cos\theta_{e\gamma}}}
= \rho_{\rm SM}(\vec{x}) 
-\frac{1}{2}\frac{E_eE_\gamma}{256\pi^6m_\mu}2\sqrt{2}G_F R_i(\vec{x})\nonumber
\\
&\equiv \rho_{\rm SM}(\vec{x}) + d_i\,\rho_i(\vec{x})
\,,
\end{align}
where $\rho_{\rm SM}$ is the SM tree-level rate and $d_i$ is the Wilson coefficient for the operator that we have added.

Given this,  one can predict how many events will occur in each bin. Here we will simplify this prediction by approximating the 
average rate for a bin
by its value at the center.

Suppose an experiment records a total of $N_\mu$ radiative muon decays. 
%(the luminosity). 
These events populate the bins in proportion to the normalized differential rate, so the expected number of events in bin $b$ is
\begin{align}\label{eq:expcounts}
\lambda_b
(d_i)
= N_\mu\,\frac{\rho_{\rm SM}(\vec{x}_b)+d_i\,\rho_i(\vec{x}_b)}{\sum_{b'}\left[\rho_{\rm SM}(\vec{x}_{b'})+d_i\,\rho_i(\vec{x}_{b'})\right]}
\,.
\end{align}

A difference in the actual number of  events and the expectation from the Standard Model (i.e. $\lambda_b
(d_i=0)$) is the signal of new physics. However, there can be statistical fluctuations around the SM value which can obscure the signal. To estimate this statistical error, we treat each bin as an independent counting experiment. 

In each bin, we get $\lambda_b$ events. 
If the true value was the SM value, the probability of obtaining this number of events is the Poisson probability
\begin{align}\label{eq:likeprod}
\cL\left(\lambda_b; d_i=0\right) = 
\frac{\left[\lambda_b(d_i=0)\right]^{\lambda_b(d_i)}}{\lambda_b(d_i)!}\,e^{-\lambda_b(d_i=0)}
\,,
\end{align}

We can compare this  to the probability of obtaining this number of events for the true value of $d_i$
\begin{align}\label{eq:likeprod}
\cL\left(\lambda_b; d_i\right) = 
\frac{\left[\lambda_b(d_i)\right]^{\lambda_b(d_i)}}{\lambda_b(d_i)!}\,e^{-\lambda_b(d_i)}
\,,
\end{align}

We can define the total difference in log-likelihood over all the bins as
\begin{align}\label{eq:likeratio}
-\ln L
&\equiv \sum_b \left|\ln \left(\frac{\cL\left(\lambda_b; d_i=0\right) }{\cL\left(\lambda_b; d_i\right) }\right)\right|
\nonumber\\
&=\sum_b |\lambda_b(d_i)\left(\ln \lambda_b(d_i=0)-\ln\lambda_b(d_i)\right)-( \lambda_b(d_i=0)-\lambda_b(d_i))|
\,,
\end{align}
 
\subsection{Numerical results}
We perform a numerical analysis with $N=5$, corresponding to a resolution of 20\% in each of the observables $(E_e/m_\mu,E_\gamma/m_\mu,\cos\theta_{e\gamma})$.
We 
impose the cuts $E_e>20~\text{MeV}$ and $E_\gamma>20~\text{MeV}$ to model a realistic detector measurement and remove the soft region. We take the number of muon events to be $N_\mu=10^5$.

\begin{figure}[t]
  \centering
  \includegraphics[width=.7\linewidth]{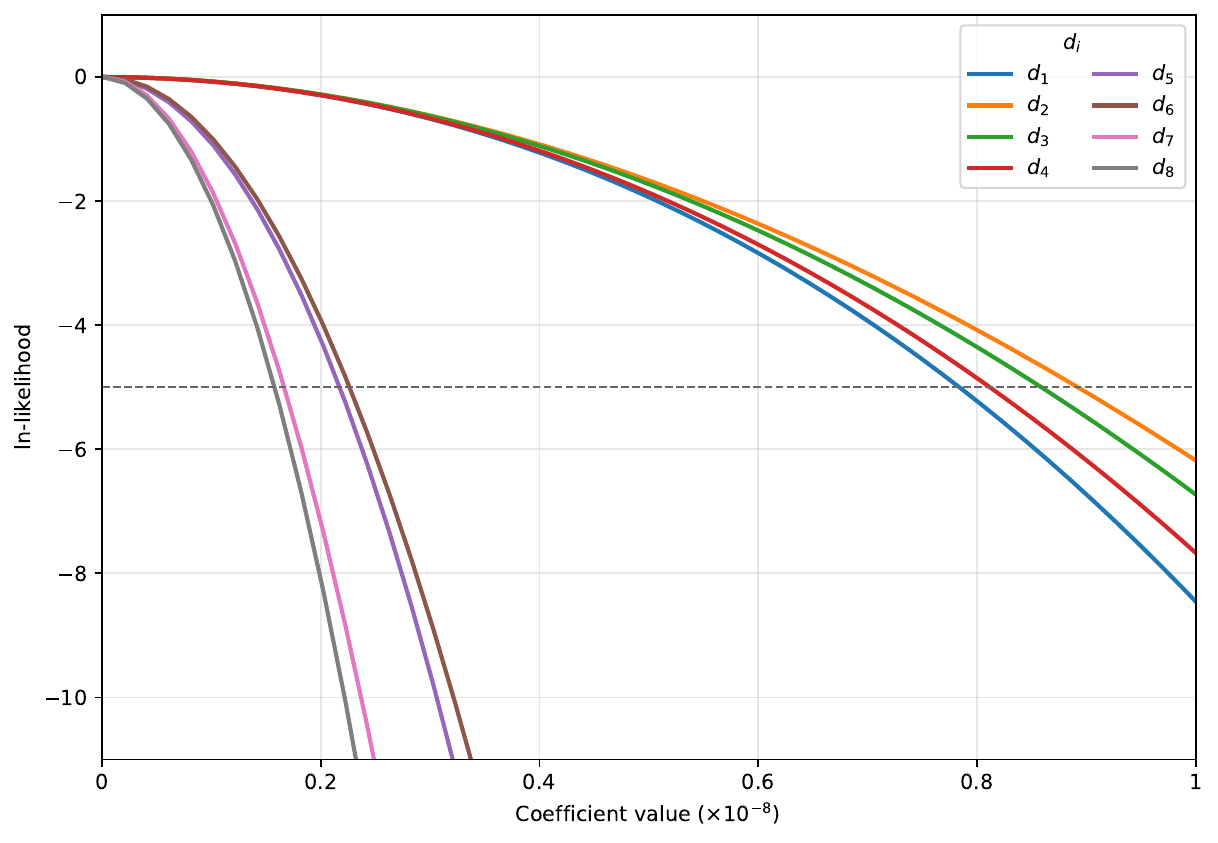}
  \caption{\small Log-likelihood for positive values of $Im\   d_i$, with $i=1,\cdots,8$.}
  \label{fig:poiratioim}
\end{figure}
\begin{figure}[t]
  \centering
  \includegraphics[width=.7\linewidth]{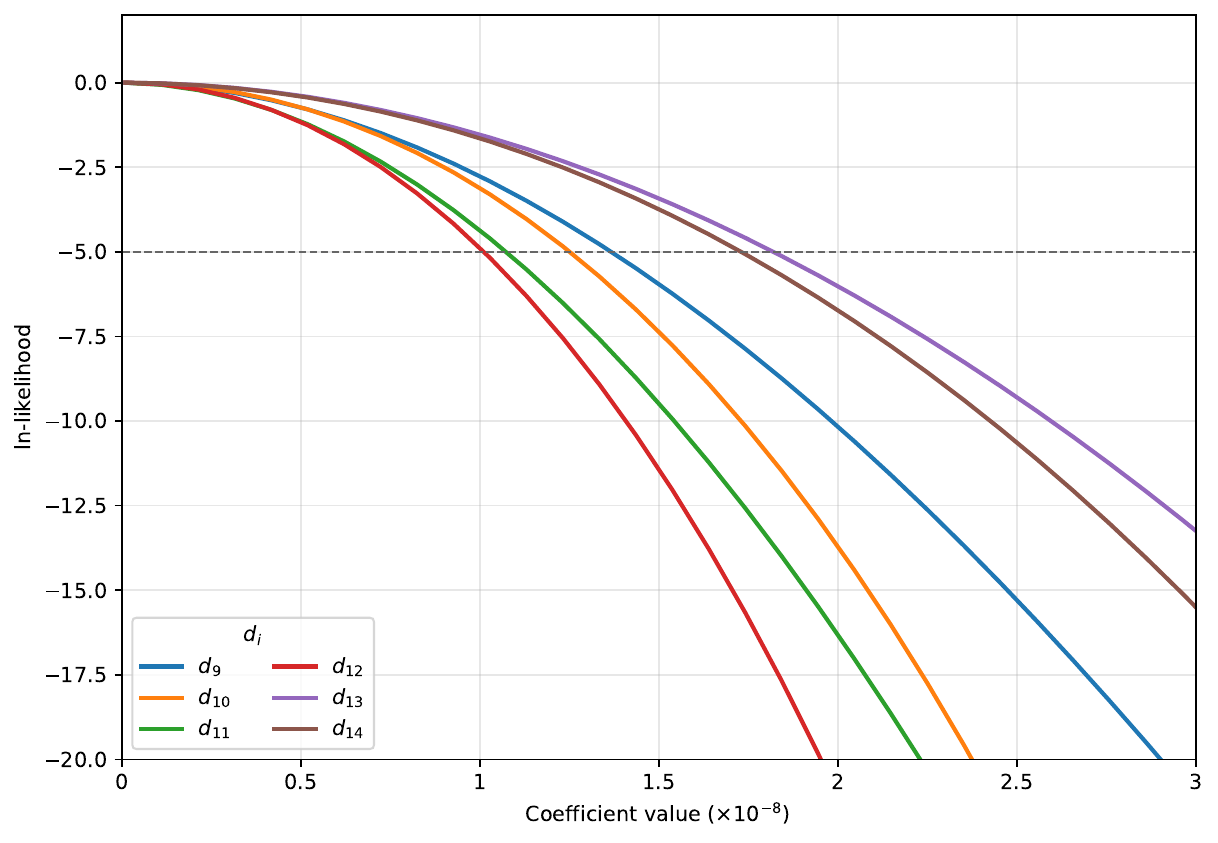}
  \caption{\small Log-likelihood for positive values of $Re\ d_i$, with $i=9,\cdots,14$.}
  \label{fig:poiratiore}
\end{figure}

\Cref{fig:poiratiore,fig:poiratioim} show the log-likelihoods  from all physical cells determined by \cref{eq:physgrid}. The first figure scans the imaginary  components of $Im\ d_{1}$ through $Im\ d_{8}$, whereas the second scans over the real components $Re\ d_9$ through $Re\ d_{14}$.

To present a more numerical estimate of the reach, we 
 consider a threshold of discovery of   $\ln L =-5$. That is, we find  the minimum value of each coefficient such  that they would  yield a  log-likelihood of -5.
The resulting limits are shown in Table \ref{tab:nresults}. We note that the operators $d_{7,8}$ are the most tightly constrained.

\begin{table}[]
\begin{tabular}{|c|c|c|c|}
\hline
Coefficient &  Limit&Coefficient &  Limit
\\
\hline
$Im\ d_{1}$ & $0.784\times10^{-8}$   &
$Im\ d_{2}$ & $0.892\times10^{-8}$    \\
$Im\ d_{3}$ & $0.859\times10^{-8}$   &
$Im\ d_{4}$ & $0.812\times10^{-8}$ 
\\
$Im\ d_{5}$ & $0.217\times10^{-8}$   &
$Im\ d_{6}$ & $0.226\times10^{-8}$  
\\
$Im\ d_{7}$ & $0.166\times10^{-8}$   &
$Im\ d_{8}$ & $0.158\times10^{-8}$  
\\
$Re\ d_{9}$ & $1.366\times10^{-8}$   &
$Re\ d_{10}$ & $1.250\times10^{-8}$ 
\\
$Re\ d_{11}$ & $1.072\times10^{-8}$   &
$Re\ d_{12}$ & $1.010\times10^{-8}$  
\\
$Re\ d_{13}$ & $1.819\times10^{-8}$   &
$Re\ d_{14}$ & $1.730\times10^{-8}$ 
\\
\hline
\end{tabular}
\caption{Values of $d_i$ for $\ln L=-5$}\label{tab:nresults}
\end{table}

While our analysis has been done for 
illustrative choices for 
the resolutions  and the cuts, it is straightforward to use the cross-section calculations for any other choice of parameters. These cross-sections can also serve as a guide to future experimental designs.

One natural extension of this work is to apply the same electromagnetic-covariant operator analysis to muon decay in orbit (DIO). In a muonic atom, the nuclear Coulomb field provides a background through which the covariant-derivative and field-strength structures considered here can contribute, potentially producing characteristic distortions of the DIO electron spectrum. Combining such an extension with established calculations of the bound-state spectrum, radiative corrections, and isotope dependence \cite{Czarnecki:2011mx,Szafron:2015DIO,Heeck:2021DIO,Szafron:2015wbm,Czarnecki:2016xbs,Fontes:2024yvw,Fontes:2025mps} could provide complementary tests of these interactions in muon-to-electron conversion experiments such as Mu2e \cite{Mu2e:2022RunI} and COMET \cite{COMET:2018TDR}.

\section*{Acknowledgements}
This work was supported in part by the US National Science Foundation under grant PHY-2210283 and a KIAS Individual Grant QP111601 via the Quantum Universe Center at Korea Institute for Advanced Study.

\bibliographystyle{apsrev4-1}
\bibliography{muon_final.bbl}

\end{document}